\documentclass{article}
\usepackage{authblk}
\usepackage{geometry}
\usepackage{amsmath}
\usepackage{graphicx} 
\usepackage[T1]{fontenc}
\usepackage{amsfonts}
\usepackage[usenames,dvipsnames]{xcolor}

\usepackage{epstopdf}%This line makes .eps figures into .pdf - please comment out if not required.

\newcommand{\be}{\begin{equation}}
\newcommand{\en}{\end{equation}}

\def\lv{{\bf l}}

\def\wv{{\bf w}}
\def\V{{\mathcal V}}

\def\R{\mathbb{R}}
\def\lin{{\rm Lin(\V)}}
\def\pt{{\partial}}
\def\ev{{\bf e}}
\def\ricci{\boldsymbol{\varepsilon}}

\def\nv{{\bf n}}

\def\tv{{\bf t}}

\def\rv{{\bf r}}

\def\tv{\mathbf{t}}

\def\uv{{\bf u}}
\def\vv{{\bf v}}
\def\kv{{\bf k}}
\def\gv{{\bf g}}
\def\Av{{\bf A}}

\def\Bv{{\bf B}}

\def\Gv{{\bf G}}
\def\Lv{{\bf L}}

\def\bom{\bm{\omega}}

\def\Sv{{\bf S}}
\def\Tv{{\bf T}}

\def\om{\boldsymbol{\omega}_s}
\def\bsi{\boldsymbol{\sigma}}
\def\Pv{{\bf P}}
\def\Iv{{\bf I}}

\def\point{p}
\def\eps{ \varepsilon}
\def\d{{\rm d}}

\def\tr{{\rm tr}}

\def\grad{\nabla}
\def\grads{\nabla\hspace{-1mm}_s}

\def\rots{\mathrm{curl}_s}

\def\dv{{\rm div}}

\def\dvs{\dv \hspace{-0.5mm}_s}
\def\esse{{{S}}}
\def\tang{{\mathcal{T}}}

\newcommand{\ot}{\otimes} 
\def\bom{\boldsymbol{\omega}}
\def\bnu{\boldsymbol{\nu}}
\def\H{{H}}

\begin{document}

\oddsidemargin = 31pt
\topmargin = 20pt

\title{Influence of the Extrinsic Curvature on 2D Nematic Films}
\author[1]{G. Napoli}
\author[2]{L. Vergori}
\affil[1]{\small Dipartimento di Matematica e Fisica ``E. De Giorgi'', Universit\`a del Salento,  73100 Lecce, Italy. E-mail: gaetano.napoli@unisalento.it}
\affil[2]{\small Dipartimento di Ingegneria, Universit\`a degli Studi di Perugia, 06125 Perugia, Italy. E-mail: luigi.vergori@unipg.it}

\maketitle
\begin{abstract}

Nematic films are thin fluid structures, ideally two-dimensional, endowed with an in-plane degenerate nematic order. In this paper we examine a generalisation of the classical Plateau problem to an axisymmetric nematic film bounded by two coaxial parallel rings. At equilibrium,  the shape of the nematic film  results from the competition between surface tension, which favours the minimization of the  area, and  the nematic elasticity which instead promotes the alignment of the molecules along a common direction.  We find two classes of equilibrium solutions in which the molecules are uniformly aligned along the meridians or parallels.  Depending on two dimensionless parameters, one related to the geometry of the film and the other to the constitutive moduli,  the Gaussian curvature of the equilibrium shape may be everywhere negative, vanishing or positive. The stability of these equilibrium configurations is investigated.

\end{abstract}

\maketitle

%\tableofcontents

\section{Introduction}
\label{}
Fluid films, such as soap films or lipid membranes,  often give rise to shapes, as beautiful as complex, that have fascinated scientists of all times and of several areas of Science. Since, as known,  the energy of an idealized two-dimensional fluid film is proportional to the area it occupies, with the surface tension being the constant of proportionality, the  minimizers  of the energy and area functionals are the same. For this reason the problem of determining  minimal surfaces with given boundaries  (raised first by Euler) has relevance not only in geometry but also in physics and engineering.  As a classical example,  a soap film attached to two twin coaxial parallel rings takes the shape of a catenoid, the only non-planar minimal surface of revolution.
On the other hand, the study of ultra thin structures subjected to the simultaneous action of various forces  gives rise to new Plateau-like problems whose solutions, besides being  of  interest from the mathematical-physic point of view,  may be used to engineer new devices controlling the geometric properties of soft shells.

An insightful approach to study the interplay between orientational order and geometry is given by  { nematic films}. These are fluid films endowed with an in-plane nematic order provided by elongated  molecules which may freely glide and/or rotate  while keeping their axes lying on the local tangent plane. 
The recent review by  Zhang {\it et al.} \cite{Zhang:2013} reports how liquid crystalline vesicles exhibit a large variety of shapes  due to the interplay  between in-plane liquid crystalline order and bending elasticity. 
 Chen and Kamien \cite{chen:2009} found axisymmetric equilibrium shapes of nematic films by minimizing a combination of surface tension and nematic elastic energies. They showed that  the nematic order is able to support a rich class of shapes in addition to the classical constant mean curvature surfaces. In the same energetic framework, Giomi \cite{giomi:2012}  searched for axisymmetric interfaces whose boundaries are two given coaxial rims and argued that only two branches of solution are allowed: the {\it catenoidal} shape when the surface tension is the dominant effect, and  the {\it pseudospherical hyperboloidal} shape when the nematic elasticity plays a predominant role. In \cite{giomi:2012} it has been  shown that the competition between nematic elasticity and surface tension induces a first order phase transition between the two branches. More recently, the same problem has been re-examined  in terms of forces by Barrientos {\it et al.} \cite{Barrientos:2017}.  
 
 It ought to be said that in all the studies quoted above only the contribution  due to the   {\it intrinsic curvatures} of the flux lines of the director field  has been accounted for in the elastic free energy of the nematics. Such a contribution   is related to the spatial variations of the director field on the curved substrate.  More  recently, it has been demonstrated that also the {\it extrinsic} curvature terms, i.e. curvatures related to the geometry of the substrate itself, are relevant in the energetic balance \cite{nave:2012,naveprl:2012, Mbanga:2012, Nguyen:2013}. The potential applications of these new theories in soft matter and their elegant mathematical formalism have produced a vivid research activity in the communities of both theoretical  physicists \cite{Jesenek:2015, Koning:2016, Gaididei:2017, Mesarec:2017, Duan:2017} and applied mathematicians \cite{Rosso:2012, segatti:2014, nave:2016, segatti:2016,  Nestler:2017}.

In this paper,  we revise the variational problem  studied in \cite{giomi:2012} in the light of the correction to the two-dimensional nematic free energy proposed in a previous work of ours \cite{naveprl:2012}. This correction   includes terms accounting for the { extrinsic} curvature of the nematic film which are instead missing in  \cite{giomi:2012}. As a result of the competition between the nematic elasticity and the surface tension,  equilibrium shapes with  positive, vanishing or negative Gaussian curvature can be obtained depending on the magnitudes of the constitutive parameters, the radius of the bounding rings and the distance between them.  The inclusion of the extrinsic curvature terms in the energy functional, on the one hand,  makes  the solutions obtained by Giomi \cite{giomi:2012} no longer admissible, on the other hand, it opens to new scenarios in which the boundary anchoring is crucial  in the determination of the equilibrium shapes. 

  {The paper is organized as follows. In Sec. II we introduce the model for the energy functional of a 2D nematic film,  and write down the equilibrium equations and  appropriate sets of boundary conditions. The specialization of the equilibrium problem to axisymmetric shapes and homogeneous alignments of the molecules of the nematics  is considered in Sec. III, where, depending on the (uniform) alignment of the director field,   the equilibrium equations are solved numerically or analytically. In Sec. IV, we study the local stability of the solutions considering both {\it in-plane strong anchoring} and {\it natural} boundary conditions on the nematic director. Sec. V contains instead some concluding remarks. The paper is closed by two appendices in which we illustrate the notation adopted throughout the paper,  report lengthy calculations and derive rigorously the equilibrium equations.}

\section{The model}
 We assume that the nematic film  is schematised by a regular  surface $\esse$ with unit normal field  $\bnu$.  We denote $\nv$  the nematic director and assume it to be  a smooth unit vector field tangent to $\esse$. The interplay between the  geometry of the film and the director field   will be studied  by minimising   the following energy functional
\be\label{energia}
W=\int_\esse\left(\gamma +  \frac{k}{2}|\grads\nv|^2\right)\d A,
\en
where $\gamma$ is  the surface tension, $k$  the elastic constant of the nematics  and $\grads$ indicates the surface gradient. 
We anticipate from the beginning that the choice of the  differential operator $\grads$ strongly affects the shape of the equilibrium  configuration. In most of the existing literature on nematic shells or films, the energy formula is usually expressed in terms of the covariant derivative  (commonly denoted ${\rm D}$) instead of the surface gradient.   What should be the most appropriate form of the energy is still on debate. In favour of  our constitutive model for the free energy it must be said that   the energy formula \eqref{energia}  can be  derived  from the  classical three-dimensional Frank's model by means of a perturbation analysis. Specifically, regarding the nematic film as a thin fluid layer whose thickness is much smaller than the minimum radius of curvature of $\esse$, to leading order  the  Frank free energy density approximates to
\be\label{ozf}
2w_{F}=k_1(\dvs\nv)^2+k_2(\nv\cdot\rots\nv)^2+k_3|\nv\times\rots\nv|^2,
\en
 where $\dvs$ and $\rots$ are, respectively, the surface divergence and surface curl \cite{nave:2012}. Next, note that, under  the one--constant approximation $(k_1=k_2=k_3=k)$, \eqref{ozf} reduces to $({k}/{2})|\grads \nv|^2$. To appreciate the differences between our model and that  studied by Giomi \cite{giomi:2012}, observe that the surface gradient and the covariant derivative of the   {director field  $\nv$ are related through the simple relation $\grads \nv = {\rm D \nv} + \bnu \otimes \Lv\nv$,  where $\Lv$ is the extrinsic curvature tensor of $\esse$. Consequently,  we have  $|\grads \nv |^2 = |{\rm D \nv} |^2 + |\Lv \nv|^2$. It is then} evident that our free energy density  exhibits an extra term reflecting the coupling of the extrinsic curvature of the film with the nematic order.

\subsection{Equilibrium equations}
The Euler-Lagrange equations associated with the energy functional \eqref{energia} can be readily derived by following consolidated variational schemes \cite{napoli:2010}.  
  {Specifically, denoting  $\bsi$, $\Tv$ and $\Gv$ the stress, couple-stress and micro-torque tensors, respectively, and $\gv$ and the micro-couple density acting on the nematic molecules,  the balance equations of forces, and macro- and micro-torques read 
\begin{subequations}
\be
\dvs \bsi = {\bf 0}, 
\label{me}
\en
\be
\dvs \Tv - \boldsymbol{\eps} \bsi = {\bf 0},
\label{couple}
\en
\be
\tv \cdot (\dvs \Gv - \gv) = 0,
\label{mstress}
\en
\end{subequations}
where  $\tv\equiv\bnu\times\nv$ represents the {\it conormal} vector, and  $\boldsymbol{\eps}$ is the Ricci alternator. Specializing the analytical scheme introduced in \cite{napoli:2010} to the energy functional \eqref{energia} gives
\begin{subequations}\label{tenvec}
\begin{align}\label{stresst}
\bsi&=\left(\gamma+\frac k2|\grads\nv|^2\right)\Pv\\
\nonumber
&-k[(\grads\nv)^T\grads\nv+(\bnu\cdot\Delta_s \nv)\bnu\ot\nv],
\end{align}
\be
\mathbf{T}=k\left[\bnu\otimes(\grads\nv)^T\tv-\tv\otimes\Lv\nv\right],
\label{stressc}
\en
\be
\Gv = k \grads \nv, \quad \gv=\mathbf{0},
\label{micros}
\en
\end{subequations}
where $\Pv \equiv \Iv - \bnu \ot \bnu $ denotes the projection onto the tangent plane,  $\otimes$ the tensor product, and $\Delta_s \equiv \dvs \grads$ is the Laplace-Beltrami differential operator on $\esse$. 

Observe first that in view of  \eqref{tenvec} and \eqref{mstress} the balance equation  of macro-moments
 \eqref{couple} is identically satisfied (see Appendix \ref{ab} for details).  Next, inserting \eqref{micros} into \eqref{mstress} yields the equilibrium equation for the in-plane orientation of $\nv$   in the simple form 
\be
\tv \cdot \Delta_s\nv=0.
\label{direttore}
\en
Adopting the most common terminology in the literature, we shall refer to \eqref{direttore} as  the {\it director equation}.
We now  denote $\ev_1$ and $\ev_2$  the principal directions on $\esse$, and parametrize  the director  through the convex angle $\alpha$ contained between $\ev_1$ and $\nv$ as 
\be\label{decomposition}
\nv= \cos \alpha \ev_1 + \sin \alpha \ev_2.
\en
 In this way, the director equation \eqref{direttore} can be rewritten as 
\be\label{director}
\Delta_s\alpha-\dvs\bom { + 2H\tau_\nv}=0,
\en
where $\bom$ is the vector parametrising the spin connection on $\esse$, $H$ is the mean curvature, and $\tau_\nv$  is the {\it geodesic torsion} \cite{docarmo} of the flux lines of the director field.

By substituting  \eqref{stresst} into \eqref{me} and projecting along $\bnu$,  we arrive at the so-called {\it shape equation} 
\begin{align}\label{shape_eq}
\nonumber
2H\left(\gamma+\frac k2|\grads\nv|^2\right)-k\Big\{&(\grads\nv)^T(\grads\nv)\cdot\Lv\\
&+\dvs[(\bnu\cdot\Delta_s\nv)\nv]\Big\}=0.
\end{align} }
  {On the other hand, the projection of  \eqref{stresst} onto the tangent plane yields (see Appendix \ref{ab}) 
\be\label{pf}
(\Delta_s\alpha-\dvs\om+2H\tau_\nv)(\grads\alpha-\om)=\mathbf{0},
\en
which, as an immediate consequence of the director equation \eqref{director},  is identically satisfied. }

\subsection{Boundary conditions}
For 2D nematic  films with boundary, the shape and director equations must be supplemented by appropriate boundary conditions. Here, we shall assume that the nematic film is {\it simply supported}, that is   { the boundary  is fixed, while the surface can freely rotate about the tangent to the boundary. This obviously entails that the component of the macro-torque along the unit tangent vector field $\lv$ must vanish, \emph{viz}, orienting $\lv$  such that $\bnu\times \lv$ coincides with the in-plane outward normal $\kv$ to $\pt S$, 
\be
\lv \cdot \Tv \kv = 0 \quad \textrm{on }\pt S.
\label{vanishingcouple}
\en
}

 As the boundary conditions on the director are concerned,  we shall take into consideration the following two cases: 
  {
 \begin{itemize}
 \item[(\emph{i})]{ {\it natural boundary conditions} which are valid whenever the molecules of the nematics may freely rotate about the normal $\nu$ at the boundary, and impose the following restriction on the micro-torque $\Gv$ \footnote{{See the first line of formula (20) in \cite{napoli:2010}. Employing the divergence theorem, one can easily deduce that the vector multiplied by the virtual rotation  $\delta \theta$}, $\Gv^T \tv$, must have zero component along the in-plane normal to $\pt S$ to allow free rotations of the molecules of the nematics at the boundary.}:
 \be
 \tv \cdot \Gv \kv = 0 \quad \textrm{on } \pt S.
 \en
} %,  and is valid whenever only degenerate anchoring is taken into account at the boundaries.}.
 \item[(\emph{ii})] {{\it in-plane strong anchoring} boundary conditions  which are valid whenever the in-plane direction of $\nv$ is fixed at the boundary. In other words, whenever the angle $\alpha$ is prescribed at the boundary.}
\end{itemize}}

\subsection{Axisymmetric shapes}
\begin{figure}[h]
\centering
\includegraphics[width=7cm,keepaspectratio]{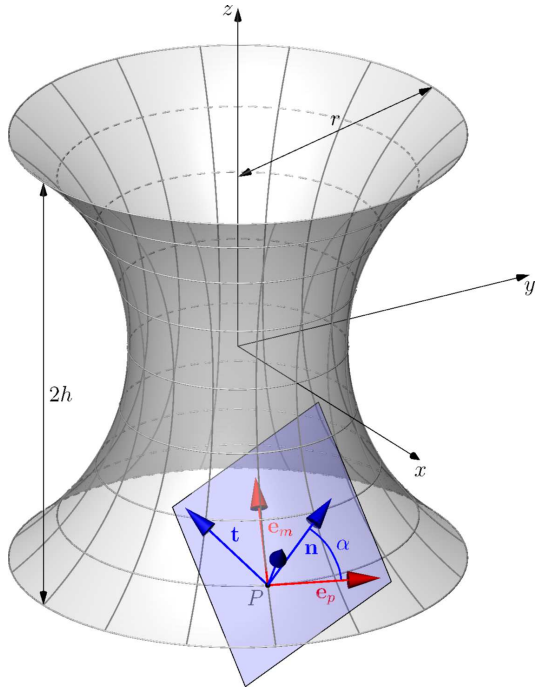}
\caption{\label{schema} Schematic representation of an axisymmetric 2D nematic film. At any point $P$ we define both the Darboux frame $\{\nv,\tv,\bnu\}$ and the orthonormal basis $\{\ev_p,\ev_m,\bnu\}$, with $\ev_p$ and $\ev_m$  being the principal directions. }
\end{figure}

Hereinafter, we shall limit our analysis to 2D nematic films schematised by axisymmetric surfaces of {\it genus }1, bounded by two fixed coaxial circular  rings of radius $r$ placed at distance $2h$ one each other as displayed in Figure \ref{schema}.   For this class of surfaces the parallels and meridians (with tangent directions $\ev_p$ and $\ev_m$, respectively) are lines of curvature, namely  $\ev_p$ and $\ev_m$ are principal directions, and the vector field $\bom$ is  divergence-free and tangent to the boundary \cite{NaVe:soft}.   {Observe now  that, according to the convention on the orientation of the unit tangent vector field $\lv$ agreed in the previous subsection, at the upper (resp. lower) boundary $\lv \equiv \ev_p$ (resp. $\lv\equiv-\ev_p$) and $\kv \equiv  \ev_m$ (resp. $\kv \equiv  -\ev_m$)}. Thus, for 2D axisymmetric films, in view of \eqref{stressc}, the boundary condition \eqref{vanishingcouple}  reads
\begin{equation}\label{bcgen}
k(\Lv\nv\cdot\ev_m)(\tv\cdot\ev_p)=0   \quad \textrm{on } \partial \esse. % \quad k\grads\alpha\cdot\ev_m=0
\end{equation}

For the sake of simplicity, we limit further our analysis to  {uniform} equilibrium alignments, i.e. homogeneous solutions to the director equation. 
Within this ansatz, equation \eqref{director} reduces to 
\be
H \tau_\nv=0,
\en
that is satisfied on the catenoid (the only surface of revolution bounded by the two given coaxial rings with vanishing mean curvature) irrespective of the (uniform) alignment of the molecules, or when the alignment of the director field is such that the geodesic torsion $\tau_\nv$ vanishes identically. But, the catenoid   satisfies the shape equation \eqref{shape_eq} if and only  if $k=0$, that is when the functional \eqref{energia} reduces to the energy of a soap film. The  classical result on the equilibrium shape of a soap film attached to two coaxial rings with the same radius is then recovered. More interestingly, the equation 
$
\tau_\nv=0
$
  implies that the director field is aligned along a principal direction on $\esse$. This means that, on  the axisymmetric surface at hand, at equilibrium the only two  uniform  alignments are those with $\nv$ oriented along  the parallels $(\alpha\equiv\alpha_p= 0)$ or the meridians $(\alpha \equiv\alpha_m= \pi/2)$.

\section{Equilibrium shapes}\label{es}
 To determine the equilibrium shapes  of the nematic film  when the molecules are oriented along the parallels or meridians,  we  express the position vector $\rv$ using cylindrical  coordinates,
$
\rv=(\rho(z)\cos \varphi,\rho(z) \sin \varphi,z),
$  with $\rho(z)>0$ for all $z\in[-h,h]$,  $\varphi\in[0,2\pi]$, and introduce the dimensionless quantities $\varrho= \rho/r$ and $\zeta = z/h$.  Thanks to such  parametrisation and nondimensionalization the energy functional \eqref{energia} may be rewritten as
\be\label{enfun}
W=2\pi\gamma r^2\int_{-1}^1w(\varrho,\varrho',\varrho'',\alpha,\alpha')\mathrm{d}\zeta,
\en
where the prime denotes differentiation with respect to $\zeta$, and 
\begin{align}\label{enden}
\nonumber
w &= \left\{1 + c\left[\frac{\varrho^2\alpha'^2+\varrho'^2+\xi^2\cos^2\alpha}{\varrho^2(\xi^2+\varrho'^2)}+\frac{\xi^2\varrho''^2\sin^2\alpha}{(\xi^2+\varrho'^2)^3}\right]\right\}\\
&\times\varrho\sqrt{\xi^2+\varrho'^2}.
\end{align}
The dimensionless parameters $\xi\equiv h/r$ and $c \equiv k/(2\gamma r^2)$  in \eqref{enden} give, respectively, a measure of the slenderness and the ratio between the magnitudes of the surface tension and elastic stiffness of the nematic film. 
 
The shape equations \eqref{shape_eq} corresponding to the two homogeneous equilibrium alignments, $\alpha\equiv\alpha_i$ ($i=p,m$), can be written as
%can be viewed as the Euler-Lagrange equations associated with the functional \eqref{enfun} with the energy density $w$ replaced by $w_i\equiv w(\varrho,\varrho',\varrho'',\alpha_i,0)$ $(i=p,m)$:
\be
\left.\left(\frac{\d}{\d \zeta^2} \frac{\pt w}{\pt \varrho''} - \frac{\d}{\d \zeta} \frac{\pt w}{\pt \varrho'} +  \frac{\pt w}{\pt \varrho}\right)\right|_{\alpha\equiv\alpha_i} =0.
\label{shapeeq}
\en
Equation \eqref{shapeeq} with $i=m$ is a fourth order ordinary differential equation (ODE), whereas for $i=p$ \eqref{shapeeq} is a second order ODE. Since the boundary is assumed  fixed, \eqref{shapeeq} must be solved subject to the boundary conditions
\be\label{rho}
\varrho(-1)=\varrho(1)=1.
\en
%and we require that the interface at boundary is torque free 
These two boundary conditions are sufficient to determine the equilibrium shapes  when $\alpha\equiv\alpha_p$. Two more boundary conditions are instead necessary when the molecules are oriented along the meridians.  Since within the parametrisation and nondimensionalization adopted here the boundary condition  $\eqref{bcgen}$ reduces to
\be\label{tzero}
\varrho''\sin^2\alpha=0\quad \textrm{at } \zeta=\pm1,
\en 
  the two additional boundary conditions to add to \eqref{shapeeq} with $i=m$  are
\be\label{rhop}
\varrho''(-1)=\varrho''(1)=0.
\en
It is worth noting that when the molecules are oriented along the parallels the boundary conditions  $\eqref{tzero}$  are identically satisfied.

For the sequent stability analysis of the equilibria  it is convenient to specify also  the boundary conditions on the angle $\alpha$. The natural anchoring boundary conditions result in the Neumann conditions
 \be\label{neumann}
 \alpha'(\pm 1)=0,
 \en
 while    {assuming that  the molecules of the nematics are forced to align tangentially  to the delimiting rims  leads to} the Dirichlet boundary conditions
  \be\label{dirichlet}
  \alpha(\pm 1)=0.
  \en
     {Obviously, both the uniform alignments $\alpha\equiv\alpha_i$ ($i=p,m$) meet the boundary conditions \eqref{neumann}, whereas $\alpha\equiv\alpha_p$ is the only uniform equilibrium alignment which satisfies   the  boundary conditions \eqref{dirichlet}.}

\subsection{Director field aligned along the parallels}
\begin{figure}[h]
	\centering
		\includegraphics[width=8cm,keepaspectratio]{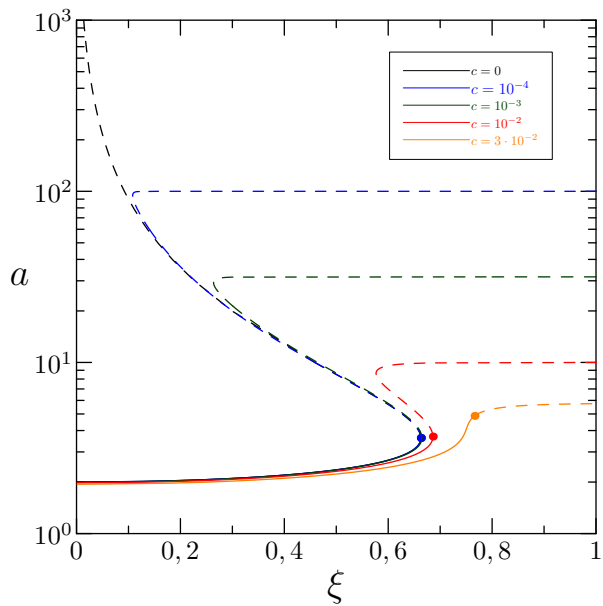}
	%	\subfigure[\label{energie}]{\includegraphics[width=6cm,height=6cm]{energie.eps}}
	\caption{Solutions to \eqref{equationa} for different values of $c$. Solid lines correspond  to stable equilibrium configurations in the case of in-plane strong anchoring (see section \ref{spa}). The dashed lines correspond instead to unstable equilibria.\label{fig:existence}}
\end{figure}

Let us now examine the equilibrium configurations  in details and start with the  case $\alpha\equiv\alpha_p$.   {In this case, the shape equation \eqref{shapeeq}} reads
\be\label{sh2}
(\varrho^2 + c) \varrho \varrho'' - ( \varrho^2 - c) ( \varrho'^2 + \xi^2) =0,
\en
and, as discussed above, has to be solved subject to the Dirichlet boundary conditions \eqref{rho}. 
The resulting boundary value problem (BVP) can be solved exactly to yield
\be\label{vpm}
\varrho_\pm (\zeta)=  \frac{\sqrt{ 2-a^2c\pm2\sqrt{1 -   a^2 c}\cosh (\xi a \zeta) }}{a},
\en
where the solution with the subscript  $+$ (respectively, $-$) refers to the case $c<1$ (respectively, $c>1$). When $c=1$ the solution of the BVP is the cylindrical shape $\varrho\equiv1$.  The positive constant $a$ in \eqref{sh2} is a root of the equation
\be\label{equationa}
\xi=\frac1a\mathrm{arccosh} \left[\pm\frac{a^2(c+1)-2}{2\sqrt{1-a^2c}}\right].
\en
For any fixed values of $c\geq c_\star\approx 0.0257$ and  $\xi>0$ equations \eqref{equationa} can be  solved uniquely for $a>0$. On the contrary, if $0 < c< c_\star$, depending on the value of $\xi$, equation \eqref{equationa} may admit one, two or three roots (see Figure \ref{fig:existence}). In the limiting case $c=0$, \eqref{equationa} admits two roots for any $\xi<\xi_\star\equiv\displaystyle\max_{s>0} \displaystyle\frac{\mathrm{arccosh} \left(s^2-1\right)}{\sqrt{2}s}\approx 0.663$, exactly one root if $\xi=\xi_\star$ and no root if $\xi>\xi_\star$. 

%Finally, for the sake of completeness, we see that also the energy of the equilibrium configurations with the molecules oriented along the parallels can be computed exactly. In terms of $a$, $\xi$ and $c$, we have
%\begin{align}\label{Wen}
%W&=2\pi\gamma r^2\Bigg\{\frac{(2+a^2c)\xi}{a}+\frac{2\sqrt{1-a^2c}}{a^2}\sinh(\xi a)\\
%\nonumber
%&+2c\:\mathrm{arctanh}\left[\frac{2(1-\sqrt{1-a^2c})-a^2c}{a^2c} \tanh\frac{\xi a}{2}\right]\Bigg\},
%\end{align}
%whence, combining \eqref{equationa} and \eqref{Wen},  the dimensionless energy $W/(2\pi\gamma r^2)$ can be specified in terms of  $\xi$ and $c$ only (Figure \ref{energie}).

\begin{figure}[h]
\centering
\includegraphics[height=8cm,keepaspectratio]{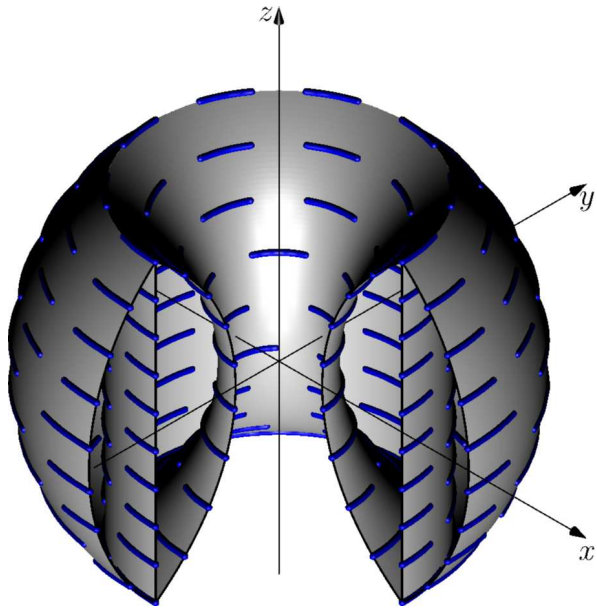}
\caption{ Equilibrium configurations with  the molecules of the nematics oriented along parallels for different values of  $c$.\label{paralleli}}
\end{figure}

   Figure \ref{paralleli} displays equilibrium shapes at different $c$.
The cylindrical configuration $(c=1)$ separates the equilibrium shape  with inward concavity ($0\leq c<1$) from those with an outward concavity ($c>1$). Consequently, at equilibrium the Gaussian curvature of the nematic film is negative if $0\leq c<1$, vanishing if $c=1$ and  positive if $c>1$. In the particular case  $c=0$, i.e. in the absence of the nematic order,   the equilibrium  shape is a catenoid.   On the contrary, in the limit as $c$ tending to infinity, that is when the effects due to the nematic elasticity are dominant,  the equilibrium  shape tends to  a portion of a sphere. This result has an intuitive explanation. At equilibrium, the  molecules of the nematics are aligned along circles whose radii is as large as possible to diminish the bending energy. This effect is in competition with the boundary conditions, which fix the radius of the boundary circles, and the surface tension, which pushes the nematic film to minimize its area and then towards the catenoidal configuration. Thus, whenever  the nematic elasticity represents the dominant contribution to the energy of the nematic film the circles far from the boundaries have larger radii, which lends the equilibrium configuration a bulgy shape. 

\subsection{Director field aligned  along the meridians}
When the molecules of the nematics are oriented along the meridians, the free energy density  depends on $\varrho''$ and, as mentioned above, the associated shape equation is a fourth-order ODE.    
The related BVP (\eqref{shapeeq}--\eqref{rhop})  can be solved only numerically. In addition, only the natural anchoring boundary conditions $\alpha'(\pm1)$ are compatible with this homogeneous alignment.

\begin{figure}[h]
\centering
\includegraphics[height=8cm,keepaspectratio]{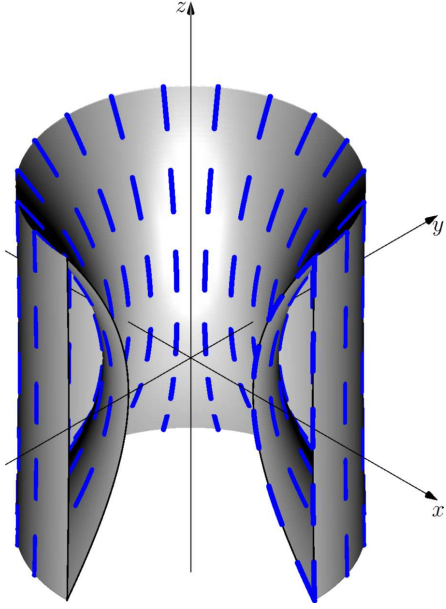}
\caption{ Equilibrium configurations with  the molecules of the nematics oriented along the meridians for different values of  $c$.\label{meridiani}}
\end{figure}

 As in the previous case,  when $c$ vanishes the equilibrium shape is  catenoidal. For greater $c$ the  nematic elasticity is more  significant and  the equilibrium shape departs from the catenoid maintaining an inward concavity (and hence the negativeness of the Gaussian curvature), though.   Also in the case $\alpha\equiv \alpha_m$ our results are in good agreement  with the physical intuition. In fact, since the flux lines of the director field are open curves (the meridians), the bending energy attains the absolute minimum when the flux lines are straight. On the other hand, the effect of the surface tension encourages the meridians to be  catenaries with inward concavity. The equilibrium configurations in Figure \ref{meridiani}  result then from the competition of these two effects. In the limiting case as $c\rightarrow +\infty$  the dominant nematic elasticity   lends the equilibrium configuration the cylindrical shape.

\section{Stability}

Let us denote $\esse_p$ and $\esse_m$ the shapes corresponding to the homogeneous  alignments $\alpha\equiv\alpha_p$ and $\alpha=\alpha_m$, respectively, and let $(\esse_p,\alpha_p)$ and $(\esse_m,\alpha_m)$ denote the two resulting equilibrium configurations. We now discuss the stability of the two equilibrium configurations under natural and  in-plane strong  anchoring boundary conditions.

 \subsection{Natural anchoring boundary conditions}
Both the classes of equilibrium shape analysed in  section \ref{es} are compatible with the natural anchoring boundary conditions.  
  The direct  calculation of the energy of the two equilibria shows that  $(\esse_m,\alpha_m)$ requires less energy than $(\esse_p,\alpha_p)$ for any values of the dimensionless parameters $c$ and $\xi$. On the other hand, the study  of the positive definiteness of the second variations at the two equilibria reveals that $(\esse_m,\alpha_m)$ is stable, that is the configuration $(\esse_m,\alpha_m)$ is a local minimizer of the energy functional \eqref{enfun}--\eqref{enden}, whereas $(\esse_p,\alpha_p)$ is unstable. 
	
	For the sake of brevity and simplicity of presentation we omit the details regarding the positive definiteness of the second variation of the energy functional at $(\esse_m,\alpha_m)$. We  instead focus on the equilibrium configuration $(\esse_p,\alpha_p)$. After some manipulations, the second variation of the energy functional \eqref{enfun}--\eqref{enden} at $(\esse_p,\alpha_p)$  can be written as
\begin{align}\label{svar}
\delta^2W[u,\vartheta]&=\underbrace{\frac{\xi^2}{2}\int_{-1}^1\left[u'^2+\frac{2\xi^2(3c-\varrho_\pm^2)}{(\varrho_\pm^2+c)^2}u^2\right]\d \zeta}_{\equiv \delta_{sh}^2W[u]}\\
\nonumber
&+\underbrace{\int_{-1}^1\frac{c\varrho_\pm}{\sqrt{\varrho_\pm'^2+\xi^2}}\left[\vartheta'^2-\frac{4c\xi^2}{(\varrho_\pm^2+c)^2}\vartheta^2\right]\d \zeta}_{\equiv\delta_{na}^2W[\vartheta]},
\end{align}
where $u\in \mathcal{H}_0^1([-1,1])\equiv\{f\in\mathcal{H}^1([-1,1]):f(\pm1)=0\}$, with $\mathcal{H}^1([-1,1])$ being  the Hilbert space of functions defined in $[-1,1]$ whose square is integrable together with the square of its weak first derivative, $\vartheta\in\mathcal{X}\equiv\{f\in\mathcal{H}^1([-1,1]):f'(\pm1)=0\}$, and $\varrho_\pm$ is given by \eqref{vpm}. {As an immediate  consequence of  \eqref{svar} we deduce that $(\esse_p,\alpha_p)$ is stable if and only if the quadratic functionals  $\delta_{sh}^2W[u]$ and $\delta_{na}^2W[\vartheta]$ are both  positive definite. But,} obviously, $\delta_{na}^2W[\vartheta_c]<0$ for any non-zero constant $\vartheta_c$. This implies that $\delta_{na}^2W[\vartheta]$ is not positive definite and thus $(\esse_p,\alpha_p)$ is an unstable equilibrium configuration.

\subsection{In-plane strong anchoring boundary conditions}\label{spa}
We now assume that the molecules of the nematics are constrained to align themselves tangentially to the boundaries. We then consider the Dirichlet boundary conditions \eqref{dirichlet}. As mentioned before, the only  homogeneous alignment compatible with these boundary conditions is that with the molecules oriented along the parallels. 

In contrast to the case of natural boundary conditions, in the case at issue the equilibrium configuration $(\esse_p,\alpha_p)$ may be stable  for some values of $c$ and $\xi$. To validate such a claim, we see that, setting 
\be
\theta=\vartheta\sqrt{\frac{\varrho_\pm}{\sqrt{\varrho_\pm'^2+\xi^2}}},
\en 
the second variation at $(\esse_p,\alpha_p)$ can be rewritten as $\delta^2W[u,\theta]=\delta^2_{sh}[u]+\delta_{sa}^2[\theta]$, with $\delta^2_{sh}[u]$ as in \eqref{svar}, 
\be
\delta_{sa}^2W[\theta]=c\int_{-1}^1\left\{\theta'^2-\frac{c[(2\varrho_\pm^2+c)(\varrho_\pm'^2+\xi^2)+\xi^2\varrho_\pm^2]}{\varrho_\pm^2(\varrho_\pm^2+c)^2}\theta^2\right\}\d \zeta,  
\en
and $u,\theta\in\mathcal{H}_0^1([-1,1])$. As {in the  case of natural anchoring boundary conditions, $(\esse_p,\alpha_p)$ is stable if and only if $\delta_{sh}^2W[u]$ and $\delta_{sa}^2W[\theta]$ are  positive-definite quadratic functionals.}

{Following standard arguments in calculus of variations, a necessary and sufficient condition for $\delta_{sh}^2W[u]$   to be positive-definite is that the interval $[-1,1]$ contains no {interior} points conjugate to $-1$ (see, for instance, \cite{gelfandfomin} page 111). For each $c>0$ we then determine the least positive value of $\xi$, say $\xi_{\rm cr}^{(sh)}(c)$, such that both the boundary value problem 
\be\label{inc1u}
\left\{\begin{array}{ll}
u''-\displaystyle\frac{2\xi^2(3c^2-\varrho_\pm^2)}{(\varrho_\pm^2+c)^2}u=0,\\
[5mm]
u(-1)=0, \quad u(1)=0,
\end{array}\right.
\en
and the normalization condition $u'(-1)=1$ (see \cite{gelfandfomin} page 106) are satisfied.  Clearly, for $\xi<\xi_{\rm cr}^{(sh)}(c)$,  the boundary value problem \eqref{inc1u}  admits only the trivial solution. Thus,  the interval $[-1,1]$ contains no interior points conjugate to $-1$ and, consequently,  $\delta_{sh}^2W[u]$ is positive-definite.}

{ Following similar arguments one can determine a necessary and sufficient condition for the positive-definiteness of $\delta_{sa}^2W[\theta]$. Specifically, denoting  $\xi_{\rm cr}^{(sa)}(c)$  the least positive value of $\xi$ for which both the boundary value problem 
\be\label{inc1red}
\left\{\begin{array}{ll}
\theta''+\displaystyle\frac{c[(2\varrho_\pm^2+c)(\varrho_\pm'^2+\xi^2)+\xi^2\varrho_\pm^2]}{\varrho_\pm^2(\varrho_\pm^2+c)^2}\theta=0,\\
[5mm]
\theta(-1)=0, \quad \theta(1)=0,
\end{array}\right.
\en
and the normalization condition $\theta'(-1)=1$ are satisfied, $\delta_{sa}^2W[\theta]$ is positive-definite if and only if $\xi<\xi_{\rm cr}^{(sa)}(c)$.}

{ We now observe that  $\delta_{sh}^2W[u]$ and $\delta_{sa}^2W[\theta]$ are both positive-definite, and hence $\delta^2W[u,\vartheta]$ is positive-definite, if and only if $\xi<\xi_{\rm cr}(c)\equiv\min\{\xi_{\rm cr}^{(sh)}(c),\xi_{\rm cr}^{(sa)}(c)\}$.}  

\begin{figure}[h]
%\psfrag{1,5}{1.5}
\centering
\includegraphics[width=10cm,keepaspectratio]{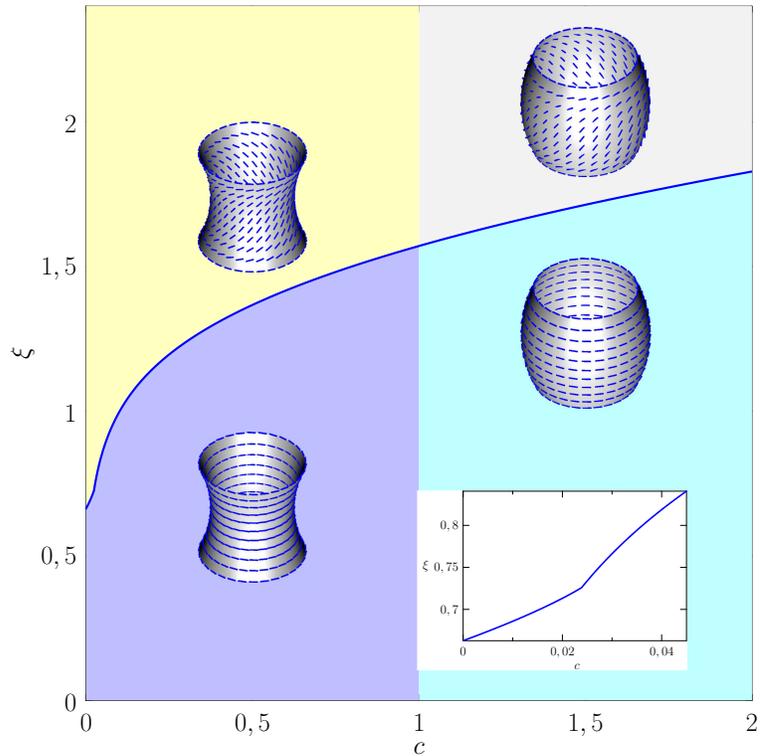}
\caption{\label{fig:xicrit} Critical threshold for the stability of $(\esse_p,\alpha_p)$ as a function of $c$.   The equilibrium configuration $(\esse_p,\alpha_p)$ is stable if and only if $\xi\leq\xi_{\rm cr}(c)$. Beyond this critical value, at a stable equilibrium configuration, the alignment of the molecules must be inhomogeneous.}
\end{figure}

 { The critical curve $\xi=\xi_{\rm cr}(c)$ displayed in Figure \ref{fig:xicrit} has been determined numerically by using Matlab \tt{bvp4c} solver}.  When the molecules of the liquid crystal are  anchored tangentially at the boundaries, the equilibrium configuration $(\esse_p,\alpha_p)$ is locally stable  if and only if $\xi<\xi_{\rm cr}(c)$.  Beyond this critical threshold,  $(\esse_p,\alpha_p)$ is no longer a local minimizer of the energy functional \eqref{enfun}--\eqref{enden}  and the equilibrium solutions bifurcate to  configurations with non-homogeneous alignments as depicted in Figure \ref{fig:xicrit}. 

\section{Conclusions}
In summary, we have investigated the equilibrium problem of fluid films endowed with  nematic order. We have showed that, as a result of the competing effects due to surface tension and orientational order, the equilibrium   shape  of the nematic film may have positive, vanishing or negative Gaussian curvature. We have presented the case of a surface bounded by two coaxial parallel rims and studied the existence, uniqueness and stability of the equilibrium configurations with  homogeneous alignments of the molecules of the nematics. Specifically, we have considered  two different sets of boundary conditions on the director field: natural and in-plane strong anchoring. In both cases we have determined locally stable equilibria, i.e. local minimizers of the energy functional.

Our analysis, though not exhaustive, shows that the inclusion of terms accounting for the extrinsic curvature  in the energy functional renders the equilibrium problem of nematic films  complex and intriguing at the same time.  Existence and uniqueness of solutions to the equilibrium equations corresponding to  non-uniform alignments of the molecules of the nematics and the search for global minimisers of the energy functional represent challenges for future analytical and numerical investigations.   Another problem worth of investigation is the generalisation of this problem  in the framework of the two-dimensional Frank's formula \eqref{ozf}, relaxing then the one constant approximation.   Motivated by the recent results by Sonnet and Virga \cite{sonnet:2017}, we think that such a generalisation leads to a more intricate scenario in the energy landscape. 

Finally, our study lays the foundations for the design of devices capable to control the shape of nematic films. To this aim, note the analogy of the nematic films studied here with the soft elastic sheets where  surfaces with both positive, vanishing or negative Gaussian curvature can be produced  by tuning the amount of local growth or swelling \cite{Klein:2007}. In the case of nematic films, an external electric or magnetic field may control the curvatures of the equilibrium shapes.

\section*{Acknowledgements}
LV gratefully acknowledges the financial support from the Italian National Group of Mathematical Physics (GNFM-INdAM) within the Young Researchers Project ``Gusci nematici sferici''. 

The authors thank  David MacTaggart for the discussions during the preparation of the manuscript.

\appendix
  {
\section{Surface differential operators} \label{app:a}

\subsection{Notation}
We first introduce the  terminology and notation adopted throughout the paper. Let $\mathcal{E}$ be a three-dimensional Euclidean point space and $\mathcal{V}$ be the Euclidean vector space associated to $\mathcal{E}$. The elements of $\V$ are three-dimensional vectors which are here denoted by lower-case boldface letters. The scalar, vector and tensor products of two vectors $\uv$ and $\vv$ are denoted $\uv \cdot \vv$, $\uv \times \vv$ and $\uv \otimes \vv$, respectively. In components, adopting the Einstein summation convention, we have $\uv \cdot \vv = u_i v_i$, $(\uv \times \vv)_i = \varepsilon_{ijk} u_j v_k$, $(\uv \otimes \vv)_{ij} = u_i v_j$, with $\varepsilon_{ijk}$ being the alternating symbol.

Second-order tensors are linear maps from $\V$ to $\V$ itself and are denoted by capital  boldface letters. The set of all second-order tensors is denoted $\lin$. The composition of two second-order tensors $\Av$, $\Bv$ is the tensor $\Av\Bv$  with components $(\Av\Bv)_{ij} = A_{ih} B_{hj}$. Once again, sum
over repeated indices is understood.  The trace is the linear operator $\mathrm{tr}:\lin\rightarrow\mathbb{R}$ which assigns to a second-order tensor $\Av$ the scalar obtained by saturation of the two indices of $\Av$, \emph{viz} $\mathrm{tr}\Av\equiv A_{ii}$. 
The superscript suffix $T$ to a second-order tensor indicates transposition: the transpose of $\Av\in\lin$ is the second-order tensor $\Av^T$ with components $(\Av^T)_{ij}=A_{ji}$. Thanks to the definitions of trace and transposition the bilinear map which assigns to two second-order tensors  $\Av$ and $\Bv$ the quantity $\Av\cdot\Bv\equiv\tr(\Av^T\Bv)=A_{ij}B_{ij}$ is a scalar product in $\lin$.

Tensors of order $n>2$ are  multilinear maps from $\V^n$ to $\mathbb{R}$. However, in this paper we consider only one tensor of order greater than 2: the Ricci alternator $\ricci$, which is a third-order tensor with components $\varepsilon_{ijk}$. Finally,   in composing tensors of different orders we agree to write the lower order
tensor on the right and  saturate all its indices. As examples of this convention,  regarding  vectors   as tensors of order one, $\Av\vv$ is a vector with components $(\Av\vv)_i= A_{ij}v_j$, $\ricci \Av$ is a vector with components $(\ricci \Av)_i = \ricci_{ijk}A_{jk}$, and $\ricci \vv$ is a second-order tensor with components $(\ricci \vv)_{ij} = \ricci_{ijk}v_{k}$.

\subsection{ Differential operators on $\esse$. The extrinsic curvature tensor.}

The nematic film is represented by a regular oriented surface $\esse$ of $\mathcal{E}$. Scalar, vector and tensor fields are functions defined on $\esse$  which assigns to each point $\point\in\esse$ an element of $\R$, $\V$ or $\lin$, respectively.
 
At each point $\point$, $\esse$ is endowed with a
2-dimensional linear space ${\tang}_\point$ called the tangent space of
$\esse$ at $\point$. The normal $\bnu(\point)$ at  $\point\in\esse$ is
one of the two unit vectors spanning  the orthogonal complement of the
tangent space. Since $S$ is orientable, at each point $p$ we can choose an orientation of the normal so that the resulting unit vector field $\bnu:S\rightarrow \V$ is differentiable.
The perpendicular projection onto the tangent plane,
$
\Pv \equiv \Iv - \bnu \ot \bnu,
$      
with $\Iv$ being the identity tensor, is then a differentiable tensor field. 

A vector field $\vv$ on $\esse$ is  tangential if $\vv(\point) \in \tang_\point$ for all $p \in \esse$. A tensor field $\Av$ on $\esse$ is tangential if, at each point $p\in\esse$, $\Av(p)\wv\in{\tang}_\point$ for all $\wv\in \V$, and $\Av(p)\bnu(p)=\mathbf{0}$.

Let $\phi$ be a differentiable scalar field on $\esse$. The surface gradient of $\phi$ is the tangential vector field
$
\grads\phi \equiv \Pv\grad\phi.
$ 
Similarly,  the surface gradient of a differentiable  vector field $\vv$ is the tensor field
$
\grads \vv\equiv (\grad \vv)\Pv.
$
The trace of $\grads\vv$ gives the surface divergence of  $\vv$, i.e. 
\be
\dvs \vv \equiv \mathrm{tr}(\grads\vv)=\grad \vv\cdot\Pv,   
\en
while twice the axial vector corresponding to the skew-symmetric part of $\grads\vv$ gives the surface curl of $\vv$, i.e.
\be
\rots \vv\equiv-\ricci\grads\vv.
\en

The tensor field $\Lv \equiv-\grads \bnu$ is  symmetric and tangential.  At each point $p\in S$, we may then regard  $\Lv(p)$ as a linear map from ${\tang}_\point$ to the tangent plane at $p$ itself whose eigenvalues $c_1$ and $c_2$ and corresponding unit eigenvectors $\ev_1$ and $\ev_2$ are the principal curvatures and directions at $p$, respectively.  The first two principal  scalar invariants of $\Lv$, 
\be\label{mean curvature}
2H  \equiv \tr (\Lv) =  - \dvs \bnu=c_1+c_2, 
\en
and
\be
\label{gauss}
K \equiv   \frac{1}{2}[(\tr\Lv)^2-\tr\Lv^2]=c_1c_2, 
\en
are the mean and Gaussian curvatures of
$\esse$, respectively. Since  $\Lv$ is a tangential tensor field, 
 the Cayley-Hamilton theorem implies that
 \begin{equation}\label{C-H}
 \Lv^2-2H\Lv+K\Pv=\mathbf{0}.
 \end{equation}
 
 Let $\nv$ be a  tangent unit vector field. The normal curvature  and the geodesic torsion along $\nv$ are defined, respectively, as
 \begin{equation}\label{curtor}
 c_\nv\equiv \nv \cdot \Lv\nv\ \quad \textrm{and}\quad \tau_\nv\equiv -\tv \cdot \Lv\nv,
 \end{equation}
 where $\tv = \bnu \times \nv$. Similarly, $c_\tv\equiv \Lv\tv\cdot\tv$ is the normal curvature along $\tv$. From this definition and \eqref{curtor} the extrinsic curvature tensor $\Lv$ can be written as
\be\label{extcurv}
\Lv=c_\nv\nv\otimes\nv-\tau_\nv(\nv\otimes\tv+\tv\otimes\nv)+c_\tv\tv\otimes\tv,
\en
 by which we readily deduce that
 \begin{equation}\label{cncth}
 c_{\mathbf{n}}+c_{\mathbf{t}}=2\H
 \quad \textrm{and} \quad
c_\nv c_\tv-\tau_\nv^2=K.
\en

 We conclude this section by reporting some identities that will be useful in deriving the equilibrium equations. Let $f$, $\uv$, $\wv$ and $\Sv$  be differentiable fields on $\esse$,  with $f$ being scalar valued, $\uv$ and $\wv$   vector valued, and $\Sv$  tensor valued. The following identities hold
\begin{subequations}\label{identitiessurf}
\be\label{idfu}
\grads(f\uv)=\uv\otimes\grads f+f\grads\uv,
\en
\be\label{idfs}
\dvs(f\Sv)=\Sv\grads f+f\dvs\Sv,
\en
\be\label{idsu}
\dvs(\mathbf{S}^T\uv)=(\dvs\mathbf{S})\cdot\uv+\mathbf{S}\cdot\grads\uv,
\en
\be\label{iduw}
\dvs(\uv\otimes\wv)=(\grads\uv)\wv+(\dvs\wv)\uv,
\en
\be\label{idurot}
(\grads\uv)\uv=\rots\uv\times\uv+\frac12\grads\left(|\uv|^2\right),
\en
\be\label{rotfu}
\rots(f\uv)=\grads f\times \uv+f\rots\uv,
\en
\be\label{idrot}
\rots\grads f=-\bnu\times\Lv\grads f.
\en 
\end{subequations}

\section{Derivation of the equilibrium equations} \label{ab}

In \cite{nave:2012} we proved that the surface gradients of the principal directions are given as follows
\begin{subequations}\label{grade}
\be
\grads\ev_1=k_1\ev_2\otimes\ev_1+k_2\ev_2\otimes\ev_2+c_1\bnu\otimes\ev_1,
\en 
\be
\grads\ev_2=-k_1\ev_1\otimes\ev_1-k_2\ev_1\otimes\ev_2+c_2\bnu\otimes\ev_2,
\en
\end{subequations}
where $k_1$ and $k_2$ are the geodesic curvatures of the curvature lines of $S$, i.e. the integral curves of the principal directions on $S$.
Then combining \eqref{decomposition},  \eqref{idfu} and \eqref{grade} the surface gradients of the director  and conormal fields are, respectively,
\begin{subequations}\label{gradnt}
\be\label{grdsna}
\grads\nv=\tv\otimes(\grads\alpha-\bom)+\bnu\otimes\Lv\nv,
\en
\be\label{gradst}
\grads\tv=-\nv\otimes(\grads\alpha-\bom)+\bnu\otimes\Lv\tv,
\en
\end{subequations}
where $\bom=-k_1\ev_1-k_2\ev_2$ is the vector parametrising the spin connection on $S$. From  \eqref{gradnt}, we readily deduce that
\begin{subequations}\label{gradsa}
\be
\rots\nv=(\grads\alpha-\bom)\times\tv-\bnu\times\Lv\nv,
\en
\be
\rots\tv=-(\grads\alpha-\bom)\times\nv-\bnu\times\Lv\tv,
\en
\end{subequations}
and
\be\label{normgn}
|\grads\nv|^2=|\grads\alpha-\bom|^2+|\Lv\nv|^2.
\en

 From \eqref{rotfu} and \eqref{grade} the surface curl of the vector parametrising the spin connection is found to be
\begin{align}\label{rotbom}
\rots\bom&=-\rots(k_1\ev_1+k_2\ev_2)\\
\nonumber
&=-(\grads k_1\times\ev_1+\grads k_2\times\ev_2)\\
\nonumber
&-|\bom|^2\bnu+k_1c_1\ev_2-k_2c_2\ev_1\\
\nonumber
&=-(\grads k_2\cdot\ev_1-\grads k_1\cdot\ev_2+|\bom|^2)\bnu\\
\nonumber
&+(\ev_2\otimes\ev_1-\ev_1\otimes\ev_2)\Lv\bom\\
\nonumber
&=-\bnu\times\Lv\bom+K\bnu,
\end{align}
where the identity
\be\label{kkk}
-(\grads k_2\cdot\ev_1-\grads k_1\cdot\ev_2+|\bom|^2)=\bnu\cdot \rots\bom=K
\en
has been used. We refer the reader to Appendix C in \cite{nave:2012} for the proof of \eqref{kkk}.
Finally, with the aid of \eqref{idrot} and \eqref{rotbom} we conclude that
\be\label{rotcov}
\rots(\grads\alpha-\bom)=-\bnu\times\Lv(\grads\alpha-\bom)-K\bnu.
\en

Next, on using    \eqref{idsu}, \eqref{C-H}, \eqref{gradnt} and the definition of the extrinsic curvature tensor we deduce that 
\begin{align}
\nv\cdot\Delta_s\nv&=\nv\cdot\dvs(\grads\nv)\\
\nonumber
&=\dvs[(\grads\nv)^T\nv]-\grads\nv\cdot \grads\nv=-|\grads\nv|^2,
\end{align}
\begin{align}
\tv\cdot\Delta_s\nv&=\tv\cdot\dvs(\grads\nv)\\
\nonumber
&=\dvs[(\grads\nv)^T\tv]-\grads\nv\cdot \grads\tv\\
\nonumber
&=\dvs(\grads\alpha-\bom)-\Lv\nv\cdot\Lv\tv\\
\nonumber
&=\Delta_s\alpha-\dvs\bom+2H\tau_\nv,
\end{align}
thanks to which the director equation \eqref{direttore} can be rewritten as \eqref{director}, and
\begin{align}\label{dnnu}
\bnu\cdot\Delta_s\nv&=\bnu\cdot\dvs(\grads\nv)\\
\nonumber
&=\dvs[(\grads\nv)^T\bnu]+\grads\nv\cdot \Lv\\
\nonumber
&=\dvs(\Lv\nv)+\Lv\tv\cdot(\grads\alpha-\bom).
\end{align}

We now report the identity
\be\label{divl}
\dvs\Lv=2\left[\grads H+(2H^2-K)\bnu\right],
\en
the proof of which is contained in Appendix A of \cite{napoli:2010}.
As a consequence  \eqref{idsu}, \eqref{gradst}, \eqref{divl} and  the symmetry of the extrinsic curvature tensor $\Lv$, we have
\begin{align}\label{dvlt1}
\dvs(\Lv\tv)&=\dvs\Lv\cdot\tv+\Lv\cdot\grads\tv\\
\nonumber
&=2\grads H\cdot\tv-\Lv\nv\cdot(\grads\alpha-\bom).
\end{align}
On the other hand, from  \eqref{extcurv},  $\eqref{cncth}_1$, \eqref{idfs}, and \eqref{gradnt} we have
\begin{align}\label{dvlt2}
\dvs(\Lv\tv)&=\dvs(-\tau_\nv\nv+c_\tv\tv)\\
\nonumber
&=-\grads\tau_\nv\cdot\nv+\grads c_\tv\cdot\tv\\
\nonumber
&-(c_\tv\nv+\tau_\nv\tv)\cdot(\grads\alpha-\bom)\\
\nonumber
&=2\grads H\cdot\tv-\grads\tau_\nv\cdot\nv-\grads c_\nv\cdot\tv\\
\nonumber
&+(\bnu\times\Lv\tv)\cdot(\grads\alpha-\bom).
\end{align}
Then, combining \eqref{dvlt1} and \eqref{dvlt2} yields
\be\label{gradcurv}
\grads\tau_\nv\cdot\nv+\grads c_\nv\cdot\tv=(\Lv\nv+\bnu\times\Lv\tv)\cdot(\grads\alpha-\bom).
\en
From  \eqref{C-H}, \eqref{extcurv}, \eqref{rotfu}, \eqref{gradsa} and \eqref{gradcurv} we obtain 
\begin{align}\label{rotln}
\rots(\Lv\nv)&=\grads c_\nv\times\nv+c_\nv\rots\nv\\
\nonumber
&-\grads\tau_\nv\times\tv-\tau_\nv\rots\tv\\
\nonumber
&=-(\grads c_\nv\cdot\tv+\grads \tau_\nv\cdot\nv)\bnu\\
\nonumber
&+(\grads\alpha-\bom)\times(\bnu\times\Lv\nv)-\bnu\times\Lv^2\nv\\
\nonumber
&=-[(\bnu\times\Lv\tv)\cdot(\grads\alpha-\bom)]\bnu-2H\bnu\times\Lv\nv+K\tv.
\end{align}

We are now in position to derive the equilibrium equations \eqref{shape_eq} and \eqref{pf}.
We first project  \eqref{me}, with $\bsi$ as in \eqref{stresst}, along the normal $\bnu$ and obtain
\begin{align}
0&=\bnu\cdot\dvs\bsi=\dvs(\bsi^T\bnu)+\bsi\cdot\Lv\\
\nonumber
&=-k\dvs[(\bnu\cdot\Delta_s\nv)\nv]+2H\left(\gamma+\frac k2|\grads\nv|^2\right)  \\
 \nonumber
&-k(\grads\nv)^T(\grads\nv)\cdot\Lv,
\end{align} 
that is equation \eqref{shape_eq}. Next, since from \eqref{mean curvature} and \eqref{iduw} one deduces that
\begin{align}
\dvs\left[\left(\gamma+\frac k2|\grads\nv|^2\right)\Pv\right]&=2H\left(\gamma+\frac k2|\grads\nv|^2\right)\bnu\\
\nonumber
&+\frac k2\grads\left(|\grads\nv|^2\right),
\end{align}
 on using \eqref{iduw}, \eqref{idurot}, \eqref{normgn}, \eqref{rotcov}, \eqref{dnnu} and \eqref{rotln},
the  projection of \eqref{me}, with $\bsi$ as in \eqref{stresst}, onto the tangent plane yields   
\begin{align}\label{eqtan}
\mathbf{0}&=\Pv\dvs\left[(\grads\nv)^T(\grads\nv)+(\bnu\cdot\Delta_s\nv)\bnu\otimes\nv\right]\\
\nonumber
&-\frac12\grads(|\grads\nv|^2)\\
\nonumber
&=\Pv\dvs\left[(\grads\alpha-\bom)\otimes(\grads\alpha-\bom)+\Lv\nv\otimes\Lv\nv\right]\\
\nonumber
&-(\bnu\cdot\Delta_s\nv)\Lv\nv-\frac12\grads(|\grads\nv|^2)\\
\nonumber
&=\Pv\Big[\rots(\grads\alpha-\bom)\times (\grads\alpha-\bom)+\rots(\Lv\nv)\times\Lv\nv\Big]\\
\nonumber
&+\frac12\grads\Big(|\grads\alpha-\bom|^2+|\Lv\nv|^2-|\grads\nv|^2\Big)\\
\nonumber
&+(\Delta_s\alpha-\dvs\bom)(\grads\alpha-\bom)+\Big[\dvs(\Lv\nv)-\bnu\cdot\Delta_s\nv\Big]\Lv\nv\\
\nonumber
&=-K\bnu\times(\grads\alpha-\bom)+(\Delta_s\alpha-\dvs\bom)(\grads\alpha-\bom)\\
\nonumber
&-[(\bnu\times\Lv\tv)\cdot(\grads\alpha-\bom)]\bnu\times\Lv\nv-[\Lv\tv\cdot(\grads\alpha-\bom)]\Lv\nv\\
\nonumber
&=(\Delta_s\alpha-\dvs\bom+2H\tau_\nv)(\grads\alpha-\bom),
\end{align}
where the last equality is a consequence of the fact that, in the light of \eqref{extcurv} and \eqref{cncth},
\begin{align}
&[(\bnu\times\Lv\tv)\cdot(\grads\alpha-\om)]\bnu\times\Lv\nv\\
\nonumber
&+[\Lv\tv\cdot(\grads\alpha-\om)]\Lv\nv\\
\nonumber
&=-[\nv\cdot(\grads\alpha-\bom)][(c_\nv+c_\tv)\tau_\nv\nv+(c_\nv c_\tv-\tau_\nv^2)\tv]\\
\nonumber
&+[\tv\cdot(\grads\alpha-\bom)][(c_\nv c_\tv-\tau_\nv^2)\nv-(c_\nv+c_\tv)\tau_\nv\tv]\\
\nonumber
&=-2H\tau_\nv(\grads\alpha-\bom)-K\bnu\times(\grads\alpha-\bom).
\end{align}
The derivation of \eqref{pf} is then complete.

Observe now that the stress tensor \eqref{stresst} is not symmetric and, in view of \eqref{dnnu} and the symmetry of $\Lv$, twice the axial vector corresponding to the skew-symmetric part of $\bsi$ is
\begin{align}\label{epssig}
\ricci \bsi&=-k(\bnu\cdot\Delta_s\nv)\tv\\
\nonumber
&=-k\Big[\dvs(\Lv\nv)+\Lv(\grads\alpha-\bom)\cdot\tv\Big]\tv.
\end{align}
On the other hand, from \eqref{iduw}, \eqref{gradnt}, \eqref{C-H}, \eqref{extcurv} and, again, the symmetry of the extrinsic curvature tensor $\Lv$, the surface divergence of the macro torque tensor \eqref{stressc} reads
\begin{align}\label{dvt}
\dvs\Tv&=-k\Lv(\grads\nv)^T\tv+k\dvs[(\grads\nv)^T\tv]\bnu\\
\nonumber
&-k(\grads \tv)\Lv\nv-k\dvs(\Lv\nv)\tv\\
\nonumber
&=-k\Lv(\grads\alpha-\bom)+k(\Delta_s\alpha-\dvs\bom)\bnu\\
\nonumber
&+k\Big[\Lv(\grads\alpha-\bom)\cdot\nv\Big]\nv-k (\Lv^2\nv\cdot\tv)\bnu-k\dvs(\Lv\nv)\tv\\
\nonumber
&=k(\Delta_s\alpha-\dvs\bom+2H\tau_\nv)\bnu\\
\nonumber
&-k\Big[\dvs(\Lv\nv)+\Lv(\grads\alpha-\bom)\cdot\tv\Big]\tv.
\end{align}
Thus, in the light of \eqref{director}, \eqref{epssig} and \eqref{dvt}  the equation of balance of macro torques is identically satisfied. }

%%%%%%%%references
%\bibliographystyle{abbrv}
%\bibliography{bibliografia_Interfaces}% Produces the bibliography via BibTeX.

\begin{thebibliography}{10}

\bibitem{Barrientos:2017}
G.~Barrientos, G.~Chac{\'o}n-Acosta, O.~Gonz{\'a}lez-Gaxiola, and J.~A.
  Santiago.
\newblock Forces on membranes with in-plane order.
\newblock {\em Journal of Physics Communications}, 1(4):045017, 2017.

\bibitem{chen:2009}
B.~G. Chen and R.~D. Kamien.
\newblock Nematic films and radially anisotropic delaunay surfaces.
\newblock {\em Eur. Phys. J. E}, 28(3):315-- 329, 2009.

\bibitem{docarmo}
M.~P. do~Carmo.
\newblock {\em Differential Geometry of Curves and Surfaces}.
\newblock Prentice-Hall, Englewood Cliffs, NJ, 1976.

\bibitem{Duan:2017}
X.~Duan and Z.~Yao.
\newblock Curvature-driven stability of defects in nematic textures over
  spherical disks.
\newblock {\em Phys. Rev. E}, 95(6):062706--, 06 2017.

\bibitem{Gaididei:2017}
Y.~Gaididei, A.~Goussev, V.~P. Kravchuk, O.~V. Pylypovskyi, J.~M. Robbins,
  D.~D. Sheka, V.~Slastikov, and S.~Vasylkevych.
\newblock Magnetization in narrow ribbons: curvature effects.
\newblock {\em Journal of Physics A: Mathematical and Theoretical},
  50(38):385401, 2017.

\bibitem{gelfandfomin}
I.~Gel'fand and S.~Fomin.
\newblock {\em Calculus of variations}.
\newblock Selected Russian publications in the mathematical sciences.
  Prentice-Hall, 1963.

\bibitem{giomi:2012}
L.~Giomi.
\newblock Hyperbolic interfaces.
\newblock {\em Phys. Rev. Lett.}, 109(13):136101--, 2012.

\bibitem{Jesenek:2015}
D.~Jesenek, S.~Kralj, R.~Rosso, and E.~G. Virga.
\newblock Defect unbinding on a toroidal nematic shell.
\newblock {\em Soft Matter}, 11(12):2434--2444, 2015.

\bibitem{Klein:2007}
Y.~Klein, E.~Efrati, and E.~Sharon.
\newblock Shaping of elastic sheets by prescription of non-euclidean metrics.
\newblock {\em Science}, 315(5815):1116--1120, 2007.

\bibitem{Koning:2016}
V.~Koning, T.~Lopez-Leon, A.~Darmon, A.~Fernandez-Nieves, and V.~Vitelli.
\newblock Spherical nematic shells with a threefold valence.
\newblock {\em Physical Review E}, 94(1):012703--, 07 2016.

\bibitem{Mbanga:2012}
B.~L. Mbanga, G.~M. Grason, and C.~D. Santangelo.
\newblock Frustrated order on extrinsic geometries.
\newblock {\em Phys. Rev. Lett.}, 108(1):017801--, 01 2012.

\bibitem{Mesarec:2017}
L.~Mesarec and W.~G. A. I.~S. Kralj.
\newblock Impact of curvature on topological defects.
\newblock {\em Journal of Physics: Conference Series}, 780(1):012015, 2017.

\bibitem{napoli:2010}
G.~Napoli and L.~Vergori.
\newblock Equilibrium of nematic vesicles.
\newblock {\em J. Phys. A: Math. Theor.}, 43(44):445207, 2010.

\bibitem{naveprl:2012}
G.~Napoli and L.~Vergori.
\newblock Extrinsic curvature effects on nematic shells.
\newblock {\em Phys. Rev. Lett.}, 108(20):207803--, 05 2012.

\bibitem{nave:2012}
G.~Napoli and L.~Vergori.
\newblock Surface free energies for nematic shells.
\newblock {\em Phys. Rev. E}, 85(6):061701--, 06 2012.

\bibitem{NaVe:soft}
G.~Napoli and L.~Vergori.
\newblock Effective free energies for cholesteric shells.
\newblock {\em Soft Matter}, 9:8378--8387, 2013.

\bibitem{nave:2016}
G.~Napoli and L.~Vergori.
\newblock Hydrodynamic theory for nematic shells: The interplay among
  curvature, flow, and alignment.
\newblock {\em Phys. Rev. E}, 94(2):020701--, 08 2016.

\bibitem{Nestler:2017}
M.~Nestler, I.~Nitschke, S.~Praetorius, and A.~Voigt.
\newblock Orientational order on surfaces: The coupling of topology, geometry,
  and dynamics.
\newblock {\em Journal of Nonlinear Science}, 2017.

\bibitem{Nguyen:2013}
T.-S. Nguyen, J.~Geng, R.~L.~B. Selinger, and J.~V. Selinger.
\newblock Nematic order on a deformable vesicle: theory and simulation.
\newblock {\em Soft Matter}, 9(34):8314--8326, 2013.

\bibitem{Rosso:2012}
R.~Rosso, E.~G. Virga, and S.~Kralj.
\newblock Parallel transport and defects on nematic shells.
\newblock {\em Continuum Mechanics and Thermodynamics}, 24(4):643--664, 2012.

\bibitem{segatti:2014}
A.~Segatti, M.~Snarski, and M.~Veneroni.
\newblock Equilibrium configurations of nematic liquid crystals on a torus.
\newblock {\em Phys. Rev. E}, 90(1):012501--, 07 2014.

\bibitem{segatti:2016}
A.~Segatti, M.~Snarski, and M.~Veneroni.
\newblock Analysis of a variational model for nematic shells.
\newblock {\em Mathematical Models and Methods in Applied Sciences},
  26(10):1865--1918, 2016.

\bibitem{sonnet:2017}
A.~M. Sonnet and E.~G. Virga.
\newblock Bistable curvature potential at hyperbolic points of nematic shells.
\newblock {\em Soft Matter}, 13:6792--6802, 2017.

\bibitem{Zhang:2013}
J.~Zhang, X.-F. Chen, H.-B. Wei, and X.-H. Wan.
\newblock Tunable assembly of amphiphilic rod-coil block copolymers in
  solution.
\newblock {\em Chem. Soc. Rev.}, 42(23):9127--9154, 2013.

\end{thebibliography}

\end{document}